\documentclass[12pt,preprint]{aastex}
 
\usepackage{graphicx}
\usepackage{epstopdf}
\usepackage{psfig}

\slugcomment{To appear in ApJ letters}
 
\shorttitle{Intensity mapping of molecular gas during cosmic reionization}
\shortauthors{Carilli}
 
\begin{document}

\title{Intensity mapping of molecular gas \\ during cosmic reionization}

\author{C.L. Carilli\altaffilmark{1}}

\email{ccarilli@aoc.nrao.edu}


\altaffiltext{1}{National Radio Astronomy Observatory, P. O. Box 0,
Socorro, NM 87801}

\begin{abstract}
\noindent 

I present a simple calculation of the expected mean CO brightness
temperature from the large scale distribution of galaxies during
cosmic reionization. The calculation is based on the cosmic star
formation rate density required to reionize, and keep ionized, the
intergalactic medium, and uses standard relationships between star
formation rate, IR luminosity, and CO luminosity derived for star
forming galaxies over a wide range in redshift. I find that the mean
CO brightness temperature resulting from the galaxies that could
reionize the Universe at $z = 8$ is $T_B \sim 1.1 (C/5)
(f_{esc}/0.1)^{-1}~\mu$K, where $f_{esc}$ is the escape fraction of
ionizing photons from the first galaxies, and $C$ is the IGM clumping
factor. Intensity mapping of the CO emission from the large scale
structure of the star forming galaxies during cosmic reionization on
scales of order $10^2$ to 10$^3$ deg$^2$, in combination with HI 21cm
imaging of the neutral IGM, will provide a comprehensive study of the
earliest epoch of galaxy formation.

\end{abstract}


\section{Introduction}

Cosmic reionization corresponds to the epoch when the neutral
intergalactic medium (IGM) is reionized by light from the first
galaxies (Loeb \& Barkana 2001).  This epoch, and the preceding 'dark
ages' between recombination and reionization, represent the 'Universum
Incognito' -- the last unexplored epoch of cosmic structure
formation. Current observational constraints, based primarily on large
scale polarization of the CMB and Ly$\alpha$ scattering of light from
$z > 6$ quasars (the Gunn-Peterson effect), suggest that reionization
may have had a large variance in space and time, starting as early as
$z \sim 15$, and extending down to $z \sim 7$ (Fan, Carilli, Keating
2006). Interestingly, a recent census of star formation in the first
galaxies suggests that, if reionization is due to the galaxies that
are seen in current deep near-IR fields (with modest extrapolation to
lower luminosity; eg. Bouwens et al. 2010), the dominant epoch of
reionization could occur at $z < 9$ (Robertson et al. 2010).

The most direct method for studying the evolution of the neutral IGM
is through the 21cm line of neutral hydrogen.  Extensive theoretical
and observational work is on-going to study the HI 21cm signal from
the neutral IGM during cosmic reionization, into the preceding
dark ages (Furlanetto et al. 2006; Morales \& Wyithe 2010).

An equally important, and complementary, area of study will be to
trace-out the large scale structure of the star forming galaxies that
reionize the Universe. Indeed, a cross correlation between the large
scale structure in star forming galaxies, and the structure of the
neutral IGM, will provide a comprehensive view of Universal evolution
during the epoch of first light. Relative to each other, these two
probes provide 'inverse views' of the Universe (Lidz et al. 2009).

The difficulty in determining the galaxy distribution is the very
large scales involved. The HI 21cm studies are probing scales ranging
from 100 deg$^2$ to 1000 deg$^2$, with depths of $z \sim 6$ to 10
(Parsons et al. 2009; Lonsdale et al. 2009, Harker et
al. 2007). Covering such large volumes with near-IR surveys of $z > 7$
galaxies is well beyond any instrumentation in the foreseeable
future. Moreover, studies of rest-frame UV light and/or the Ly$\alpha$
emission from galaxies well into reionization likely provide a biased
view of the galaxy distribution due to the combination of strong
Ly$\alpha$ absorption by the neutral IGM, and any dust in the host
galaxies (Fontana et al. 2010).

In this letter, I consider the possibility of tracing the large scale
structure of star forming galaxies during reionization using the
technique of intensity mapping of the molecular gas in star forming
galaxies, as traced via the CO emission lines. Intensity mapping
involves imaging the aggregate emission from thousands (or more) of
galaxies on very large scales (hundreds of comoving Mpc). Individual
galaxies are not detected, just the summed signal from the large scale
distribution of galaxies. This technique has been explored using the
HI 21cm line out to $z \sim 1$ (Chang et al. 2010).  Note that the
ALMA and the EVLA are able to detect the CO emission from individual,
massive galaxies beyond $z \sim 6$ (Wang et al. 2010; Carilli 2010;
Walter \& Carilli 2007), however, the field of view of these
instruments is small ($< 1'$), precluding surveys on scales of tens of
degrees.  Intensity mapping offers a straight-forward alternative for
tracing the very large structure in the galaxies that reionize the
neutral IGM, using smaller area telescopes, without the need to detect
galaxies individually.

My goal is to estimate the mean brightness of the CO emission on large
scales from the galaxies that reionize the neutral IGM.  The notion
that molecular line emission from galaxies at $z = 0$ to 10, could be
a substantial CMB foreground was considered by Righi et al. (2008)
using an extended Press-Schecter formalism for merging halos to set
the evolution of galaxies, with analytic recipes for the implied CO
emission from these galaxies. In this letter, I take a simpler
approach, based on the required cosmic star formation rate density to
reionize the Universe.  This approach allows for a straight-forward
derivation of the mean CO surface brightness from the galaxies that
could reionize the neutral IGM at a given redshift, although it does
not predict the structure in this emission.  Clearly, the
uncertainties in the assumptions are significant, and these are
pointed out below. However, the estimate is meant to be order of
magnitude, and useful for initial consideration of future
instrumentation to perform CO intensity mapping.

\section{Calculation}

Following is a calculation of the mean CO brightness expected from the
galaxies that could reionize the Universe at a given redshift. I adopt
the standard concordance cosmology (Spergel et al. 2007), and comoving
coordinates are used throughout.  The basic information that goes into
the calculation can be summarize as follows:

\begin{itemize}

\item The cosmic star formation rate density required to reionize (and
keep ionized) the neutral IGM. 

\item A conversion from star formation rate to IR luminosity based on
known properties of galaxies. 

\item A conversion from IR luminosity to CO luminosity based on known
properties of galaxies, at least out to $z \sim 3$.

\end{itemize}

\subsection{Star formation rate density and IR luminosities} 

Madau, Haardt, and Rees (1999) calculate the star formation rate
density ($\dot{\rho}_{SFR}$ in $\rm M_{\odot}~ yr^{-1}~ Mpc^{-3}$,
comoving) required to reionize the IGM. The basic assumptions are the
mean baryon density, the clumping factor (recombinations), the escape
fraction of ionizing radiation from galaxies, and a Salpeter IMF over
0.1 to 100 M$_\odot$ range. The Madau et al. calculation has been
updated to the latest cosmological parameters by Bunker et al. (2010)
in their section 3.3.  This relation is key to the calculation below,
and I repeat it here:

$$ (1) ~~~ \dot{\rho}_{SFR}^{reion} \equiv {\rm SFR/Volume} = 0.005~
f_{esc}^{-1}~ [{{(1+z)}\over8}]^3~ (C/5) X ~ \rm M_{\odot}~ yr^{-1}~
Mpc^{-3} $$

\noindent where $f_{esc}$ is the ionizing photon escape fraction, $C$
is the clumping factor, and $X$ is a factor that depends on the cosmic
baryon density and h$_o$, and equals 1 for the standard cosmology ($X$
will be removed from hereon).  Simulations suggest a value of $C$
somewhere between 5 and 30 (Furlanetto et al. 2006).  The value of
$f_{esc} \sim 0.06$ for the Milky Way and other nearby star forming
galaxies (Putman et al. 2003). Shapley et al. (2006) show that this
may rise to 0.1 to 0.2 for $z \sim 3$ LBGs (see also Nestor et
al. 2011), and there is marginal evidence that this may increase
further to higher redshift (Bouwens et al. 2009).

The relationship between far IR luminosity (FIR) and star formation
rate has been considered in numerous studies over the last two
decades. I adopt the standard relationship for nearby
galaxies given in Kennicutt (1998):

$$ (2) ~~~ L_{FIR} = 1.1 \times 10^{10} ~ \rm SFR ~ L_\odot $$

\noindent where SFR is in M$_\odot$ yr$^{-1}$.  I will assume in the
following calculation that this relationship applies on average to all
galaxies, and hence applies to the volumetric average over galaxies on
large scales.

In this case, the SFR as a function of $L_{FIR}$ can be solved for in
equation (2), and substituted into equation (1), yielding the FIR
luminosity per unit cosmic volume required to reionize the neutral IGM:

$$ (3) ~~~  \rho_{FIR} = 5.5 \times 10^{7} f_{esc}^{-1}~
[{{(1+z)}\over8}]^3~ (C/5)~ ~ \rm L_\odot~ Mpc^{-3}$$

\noindent where $\rho$ in this, and subsequent, equations indicates
the comoving volume density of the subscripted quantity, in this case,
the FIR luminosity.

\subsection{Cosmological relationships for CO and FIR luminosities}

The 'integrated Kennicutt-Schmidt law' relates the FIR luminosity of
galaxies to the CO luminosity, and also has been a topic of extensive
study for both nearby and distant galaxies.  The underlying physical
relationship is between the star formation rate and the gas mass, but
for the purposes of this letter, the critical component is the
observational relationship.

In this paper, I employ the latest empirical relationships between CO
and FIR luminosity derived by Daddi et al. (2010) for both nearby and
distant star forming galaxies.  They conclude that there are two,
roughly linear, relationships between CO luminosity and FIR luminosity
for galaxies, one relevant for normal star forming galaxies at both
low and high redshift, including the sBzK galaxies at $z \sim 2$, and
one relevant for dense starburst galaxies, such as the ULIRGs nearby,
and the submm galaxies and quasar hosts in the distant Universe. I
will use the median ratio they derive for the dominant, gas rich, star
forming disk galaxy population at low and high redshift:

$$ (4) ~~~  L'_{CO} = 0.02 L_{FIR}~\rm K~ km~ s^{-1}~ pc^2 $$

\noindent where $L_{FIR}$ is in L$_\odot$. 

The units for $L'_{CO}$ were originally designed for spatially
resolving observations of CO in the Galaxy, where brightness
temperature was paramount, and from which the empirical relationships
between molecular gas mass (H$_2$) and CO luminosity have been
calibrated.  I return to this point below.

We next consider the standard relationships between CO luminosity and
observed flux from Solomon \& Vanden Bout (2006). Since these
relations are key to the calculation, I repeat the critical relations
here.

Solomon \& Vanden Bout relate CO luminosity to flux density and line width 
as follows: 

$$ (5) ~~~ L_{CO} = 1.0\times 10^{3} S \Delta V (1+z)^{-1}\nu_r  D_L^2 ~~~ \rm L_\odot $$

\noindent where the luminosity distance, $D_L$, is in Gpc, $\nu_r$ is
the rest frequency in GHz, flux density, $S$, is in Jy, and velocity
width, $\Delta V$, is in km s$^{-1}$. They also give a related equation for
$L'_{CO}$:

$$ (6) ~~~ L'_{CO} = 3.3 \times 10^{13} S \Delta V D_L^2 \nu_o^{-2}(1+z)^{-3} \rm ~~~
K~ km~ s^{-1}~ pc^2 $$

\noindent where $\nu_o$ is the observing frequency in GHz. Solving for
$S\Delta V$ in (5) and (6), and equating, yields:

$$ (7) ~~~ L_{CO} = 3\times 10^{-11} \nu_r^3 L'_{CO} ~~~ \rm L_\odot $$

\noindent The cubic dependence on $\nu_r$ comes from the fact that
$L'_{CO}$, being in K, is independent of transition (for constant
brightness temperature, $T_B$), hence the luminosity will increase as
$\nu_r^2$ due to the definition of $T_B$, and another factor of $\nu_r$
due to the increased line width in Hz for the higher transitions, for
a fixed velocity width in km s$^{-1}$.

Substituting for $L'_{CO}$ in equation (7) using (4) then yields: 

$$ (8) ~~~ L_{CO} = 6\times 10^{-13} \nu_r^3 L_{FIR} ~~~ \rm L_\odot $$

\subsection{The CO emission from galaxies that reionize the Universe}

Again, making the assumption that the individual galaxy relationships
apply to the volumetric average of galaxies on large scales, we can
combine equations 8 and 3 to obtain the CO luminosity per unit volume
required to reionize the neutral IGM:

$$ (9) ~~~ \rho_{L_{CO}} = 6.4 \times 10^{-8} \nu_r^3 f_{esc}^{-1}
(1 + z)^3 (C/5) ~~~ \rm L_\odot~ Mpc^{-3} $$

Equation (9) is the fundamental relationship for the expected CO
luminosity per comoving cosmic volume dictated by the star formation
rate density required to reionize the neutral IGM, based on empirical
relationships between $L_{CO}$, $L_{FIR}$, and star formation rate for
star forming galaxies at low and high redshift.

We can go further, and consider observable quantities. Equation (5)
gives the relationship between observables $S$ and $\Delta V$ and
the CO luminosity for a given galaxy. Again, assuming this applies to
a volumetric average of galaxies (ie.  $\rho_{L_{CO}}$), and using
equation (9) and (5), we obtain a velocity integrated CO flux per
unit comoving volume required for reionization:

$$ (10) ~~~ \rho_{S \Delta V} = 6.4\times 10^{-11} \nu_r^2
D_A^{-2} f_{esc}^{-1} (C/5) ~~~ \rm Jy~ km~ s^{-1}~ Mpc^{-3} $$

\noindent where $D_A$ is the angular diameter distance in Gpc. 
$D_A \sim 1.0 \pm 0.2$ Gpc for $z = 6$ to 10.  

\section{Examples}

Let us use equation (10) in an example. Consider a survey at $5'$
resolution at $z = 8$ for the CO 2-1 line (230GHz rest frequency,
25.6GHz observing frequency). The angular size corresponds to 13 Mpc
(comoving), which implies a $\Delta z = 0.044$ for the standard
cosmological expansion.  The comoving cosmic volume covered is then
2200 Mpc$^3$.  Multiplying through by this volume, equation (10) then
yields $S \Delta V = 0.073 (f_{esc}/0.1)^{-1} (C/5)$ Jy km s$^{-1}$.
The $\Delta z$ due to cosmic expansion corresponds to a depth of 1500
km s$^{-1}$, yielding a mean signal over the band of: $S = 50
(f_{esc}/0.1)^{-1} (C/5)$ $\mu$Jy.
 
We can then use the standard Rayleigh-Jeans relationship to derive
observed brightness temperature: $T_B = 1360 S ~ \lambda^2
\theta^{-2}~~~ \rm K$, where $\theta$ is the angular size in
arcseconds, $S$ is the flux density in Jy, and $\lambda^2$ is the
observing wavelength in cm.  For a 50$\mu$Jy signal at $5'$ resolution
and observing at 1.2cm wavelength: $T_B = 1.1 (f_{esc}/0.1)^{-1} (C/5)
~\mu$K.

Performing the same calculation at $15'$ resolution and $\Delta z =
0.13$, yields: $S = 450 (f_{esc}/0.1)^{-1} (C/5)$ $\mu$Jy, and again,
$T_B = 1.1 (f_{esc}/0.1)^{-1} (C/5) ~\mu$K. Likewise, for CO 1-0 at
$15'$ resolution, the values are $S = 110 (f_{esc}/0.1)^{-1} (C/5)$
$\mu$Jy, and again, $1.1 (f_{esc}/0.1)^{-1} (C/5) ~\mu$K.

Based on this calculation, we can estimate the expected mean CO
brightness temperature (observed-frame) as a function of redshift, for
galaxies that can reionize the neutral IGM at a given redshift. The
result is shown in Figure 1. Note that this is independent of
transition (assuming constant $T_B$ for the low order transitions),
and independent of resolution (being a mean surface brightness).

\section{Discussion}

I have considered the requisite CO emission coming from the galaxies
that reionize the Universe at a given redshift. The calculation above
leads to a mean surface brightness at $z = 8$ of:

$$ T_B \sim 1.1 (f_{esc}/0.1)^{-1} (C/5) \rm ~~~\mu K $$

\noindent with a gradual variation with redshift (Figure 1).

Note that real observations will entail detecting fluctuations in the
CO signal around the mean.  Also, this value of $T_B$ is dictated by
the star formation rate density at a given redshift that is needed to
reionize, and keep ionized, the neutral IGM (equation (10)). In
reality, the star formation rate density will evolve as it will, and
we will observer over what redshift range the IGM reionizes.

There are clearly substantial uncertainties in the calculation above.
Explicit uncertainties involve $f_{esc}$ and $C$, both of which 
have current estimates that could range by a factor of a few (see
section 2.1). Determining the escape fraction of ionizing photons 
will be a major goal of future observations of $z > 7$ galaxies with 
JWST, while determining the clumping factor is clearly a goal of next
generation HI 21cm reionization experiments.

Implicit to the calculation are the relationships between star
formation rate, FIR luminosity, and CO luminosity. The dust and CO
production and heating in the first galaxies remains uncertain. Hence,
it remains unclear whether one can extrapolate standard
Kennicutt-Schmidt star formation laws, and FIR luminosity to star
formation rate relations, to very early galaxies.  Detection of strong
CO line emission, and thermal emission from warm dust, from $z > 6$
quasar host galaxies suggests that CO and dust can be produced rapidly
in the first star forming galaxies, although these systems are at the
high mass end of the galaxy distribution (Wang et al. 2008, 2010).  As
for CO excitation, clearly the CMB will 'depopulate' the lowest levels
at high $z$, although a calculation of the Boltzmann distribution
suggests only a factor two or so decrease in the population of the
first excited state at $z \sim 7$ in a typical star forming galaxy,
relative to having no background at all. Calibrating the relationships
between CO luminosity, FIR emission, and star formation from $z > 7$
galaxies will be a major goal with ALMA, the EVLA, and the JWST, in
the coming years, on a galaxy-by-galaxy basis. This information can
then be incorporated into the calculation above for a better estimate
of the average CO emission seen on very large scales by intensity
mapping.

Another difficulty will be line confusion. The observed line cubes
will contain different transitions at different redshifts. To make
full use of the data a dual-frequency experiment may be required,
probing two transitions of CO to eliminate ambiguities, although cross
correlation with the HI 21cm images will also help.

Overall, the estimate of $T_B$ for CO above should be considered at
best accurate to an order of magnitude. We can compare this value to
more involved calculations in the recent literature.  Righi et
al. (2008) have considered the background radiation expected from CO
lines from galaxies from $z = 1$ to 10, in the frequency range of
$\sim 10$ to 50 GHz. Their calculation adopted the extended
Press-Schecter formalism to set the galaxy merger rate, and used
analytic recipes for the implied CO emission from these galaxies. They
obtain a mean signal from a given CO transition at $z \sim 2$ within a
factor two of the mean calculated herein. More recently, Gong et
al. (2011) have used the Millennium simulation results of Obreschkow
et al (2009) to estimate the CO surface brightness fluctuations on
large scales at $z > 6$.  The Obreschkow et al. study involved
generating mock galaxy catalogs in CO emission based on the Millennium
dark matter cosmological simulation, and again, invoking recipes to
extrapolate to CO luminosity for a given halo. Gong et al. find a mean
CO surface brightness of $\sim 0.1\mu$K to $0.7~\mu$K at $z \sim 7$
(depending on model assumptions), with fluctuations a factor of a few
lower than this on scales $\sim 10'$ (see also Furlanetto et al. 2011,
in prep).  The roughly factor two agreement between these
sophisticated calculations and the simple estimate presented herein is
encouraging, given the very different approaches.

Is it plausible to detect such a signal?  Without going into detailed
instrument design, we can consider receiver noise and brightness
temperature (ie. the radiometry equation).  For a 20K receiver system
in 1000 hours and a 130 MHz channel (= 1500 km s$^{-1}$ at 26GHz):
$ \delta T = 20{\rm K} / [3.6\times 10^6 \times 1.3\times 10^8]^{1/2} = 0.9
\rm ~~~ \mu K $. Hence, achieving a $1 \sigma$ brightness sensitivity is
plausible in a long integration.

A key point to keep in mind is that the initial studies will be
statistical in nature, eg. a cross correlation with the HI 21cm
images, or study of the autocorrelation of the CO cubes themselves.
The challenge will be obtaining the combination of very wide field ($>
100$ deg$^2$) with $\mu$K surface brightness sensitivity at
resolutions $\sim 1'$ to $10'$, at frequencies between 15 and 45
GHz. Gong et al. (2011) consider this point in more detail. They
estimate that a 1000 element array of 0.7m diameter antennas with a 1
GHz bandwidth and 30 MHz channels operating at 15GHz could detect the
CO power-spectrum to high significance in 3000 hours on comoving
scales of k $\sim 0.1$ Mpc$^{-1}$ to 1 Mpc$^{-1}$. Bowman et al. 2011
(in prep) are also considering the design for such an experiment,
including the relative benefits of interferometers versus focal plane
arrays on single dish telescopes.

Ultimately, the critically unique analysis will be the cross
correlation with the HI 21cm image cubes (Gong et al. 2011; Furlanetto
et al. 2011), which is well beyond the scope of this letter. My main
conclusion is that the simple calculation presented herein, based on
the star formation rate required to reionize the Universe, supports
the more sophisticated treatments of the problem, and suggests that
the CO intensity mapping experiment at $z > 6$ is plausible.

\acknowledgements CC thanks the Keck Institute for Space Studies for
instigating this work, the CO intensity mapping design team for
lively discussions, and the referee for insightful comments that
improved this paper.

\clearpage
\newpage

\noindent Bouwens, R. et al. 2009, ApJ, 705, 936

\noindent Bouwens, R. et al. 2009, ApJ, 705, 936

\noindent Bouwens, R. et al. 2010, ApJ, 709, L133

\noindent Bunker, A. et al. 2010, MNRAS, 409, 855

\noindent Chang, Tzu-Ching, et al. 2010, Nature, 466, 463

\noindent Carilli, C.L. 2010, AIPC, 1294, 234

\noindent Daddi et al. 2010, ApJ, 714, L118

\noindent Fan, X., Carilli, C., Keating, B. 2006a, ARAA, 44, 415

\noindent Fontana, A. et al. 2010, ApJ, 725, L205

\noindent Furlanetto, S., Oh, S.P., Briggs, F. 2006, PhysRep., 
433, 181

\noindent Gong, T. et al. 2011, ApJ (letters), in press

\noindent Harker, G. et al. 2010, MNRAS, 405, 2492

\noindent Kennicutt, R. 1999, ARAA, 36, 189

\noindent Lidz, A. et al. 2009, ApJ, 690, 252

\noindent Loeb, A. \& Barkana, R. 2001, ARAA, 39, 19

\noindent Lonsdale, C.J. et al. 2009, IEEE, 97, 1497

\noindent Madau, P., Haardt, Rees, M. 1999, ApJ
514, 698

\noindent Morales, M. \& Wyithe, S. 2010, ARAA, 48, 127

\noindent Nestor, D., Shapley, A., Steidel, C., Siana, B. 2011, ApJ, 
submitted (arXiv:1102.0286v1)

\noindent Obreschkow, D. et al. 2009, 698, 1467

\noindent Parsons, A. et al. 2010, AJ, 139, 1468

\noindent Putman, M. et al. 2003, ApJ, 597, 948

\noindent Robertson, B. et al. 2010, Nature, 468, 49

\noindent Righi, M., Hernandez-Monteagudo, C., Sunyaev, R. 2008, A\& A, 
489, 489

\noindent Shapley, A. et al. ApJ, 651, 688

\noindent Solomon, P. \& vanden Bout, P. 2006, ARAA, 43, 677

\noindent Spergel, D. et al. 2007, ApJS, 179, 377

\noindent Wang, R. et al. 2010, ApJ, 714, 699

\noindent Wang, R. et al. 2008, ApJ, 687, 848

\noindent Walter, F. \& Carilli, C.L. 2007, HiA, 14, 263

\clearpage
\newpage

\begin{figure}
\psfig{file=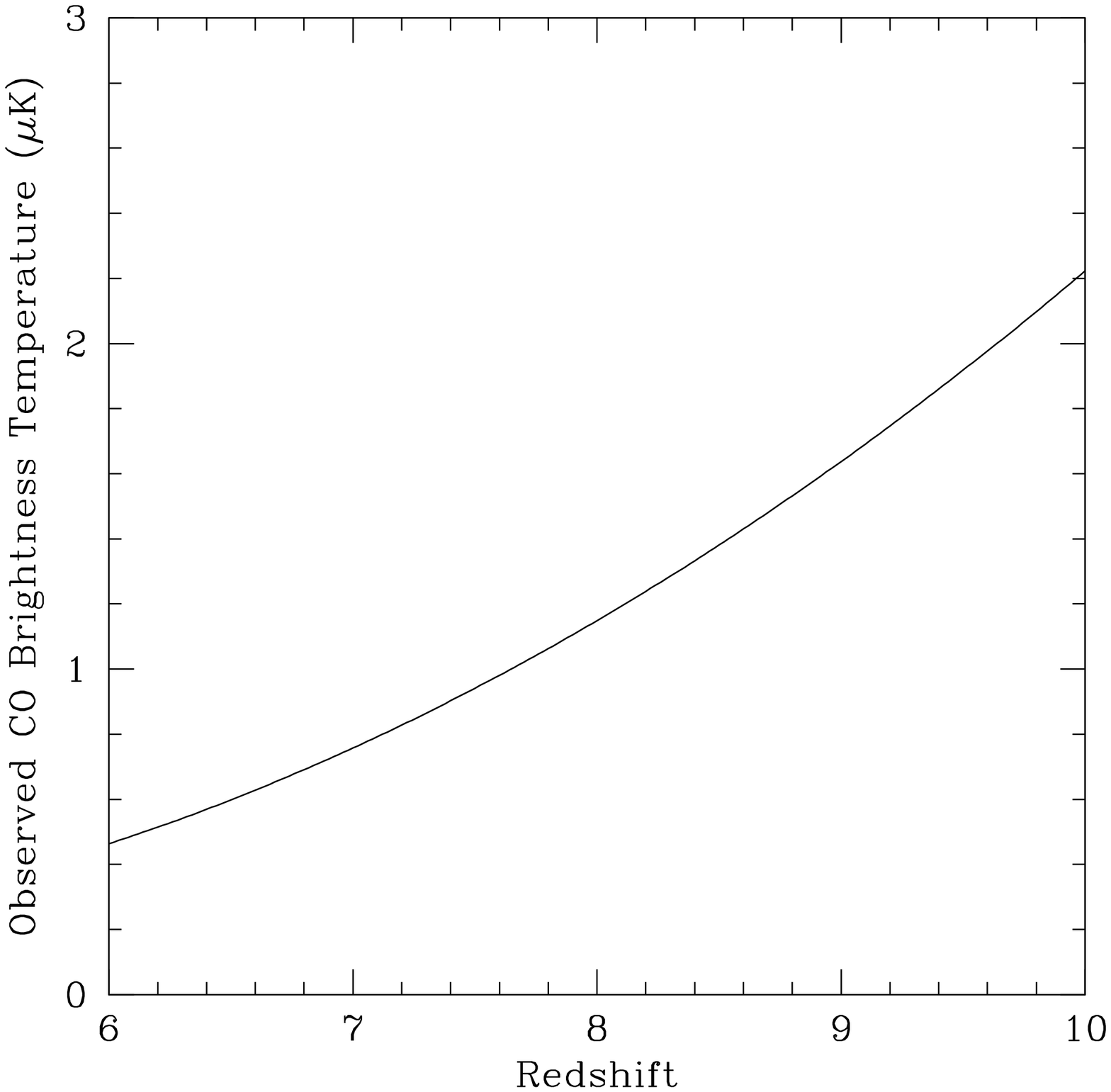,width=5in}
\caption{
The expected mean CO brightness temperature (observed-frame) 
for galaxies that can reionize, 
and keep ionized, the neutral IGM at a given redshift, 
based on equation (10) and assuming $f_{esc} = 0.1$ and $C = 5$. 
}
\end{figure}

\end{document}